\documentclass{jps-cp}
\usepackage{txfonts} 

\title{Commissioning and Performance of the New ALICE Inner Tracking System in the First Phase of LHC Run 3}

\author{Ivan \textsc{Ravasenga}$^{1}$, for the ALICE Collaboration}

\inst{$^{1}$CERN, Geneva, Switzerland}

\email{ivan.ravasenga@cern.ch}

\recdate{December 16, 2022}

\abst{ALICE is one of the four experiments at the CERN Large Hadron Collider (LHC) specifically designed to study nuclear matter at extreme conditions of temperature and pressure. The LHC Run 3 started officially in July 2022 with proton--proton (pp) collisions at a centre-of-mass energy \mbox{$\sqrt{s}$ = 13.6 TeV} after a shutdown of more than two years to allow for the upgrade of both the accelerator and the experiments. To fulfill the requirements of the ALICE physics program for Run 3, new timing, tracking and multiplicity detectors have been installed in the ALICE central barrel both at mid- and forward rapidities, the readout electronics of the majority of the detectors has been upgraded and the ALICE Offline and Online computing farm has been completely renewed. A key detector of the apparatus is the new, ultra-light, high-resolution Inner Tracking System (ITS2) that significantly enhances the resolution on the determination of the distance of closest approach to the primary vertex, the tracking efficiency
at low transverse momenta, and the read-out rate capabilities with respect to what was achieved in the LHC Run 2 with the older tracker. The new tracker consists of seven
layers, azimuthally segmented into Staves, equipped with silicon Monolithic Active Pixel Sensors with a pixel size of 27$\times$29 $\mu$m$^{2}$ covering a sensitive area of about 10 m$^2$.
A long effort allowed the construction, characterization, installation and commissioning of the ITS2 before starting Run 3. The in-pixel discriminator threshold calibration represents a challenging procedure due to the huge number of channels and it is carried out by exploiting the power of the ALICE Offline and Online farm. The detector took data for the full 2022 and its first performance results represent the second key part of this article.}

\kword{LHC, ALICE, tracker, track, cluster, pixel, silicon, run, calibration, detector, alignment}

\begin{document}
\maketitle
\section{The upgraded Inner Tracking System}\label{sec:intro}
\begin{figure}
\centering
\includegraphics[width=0.6\textwidth]{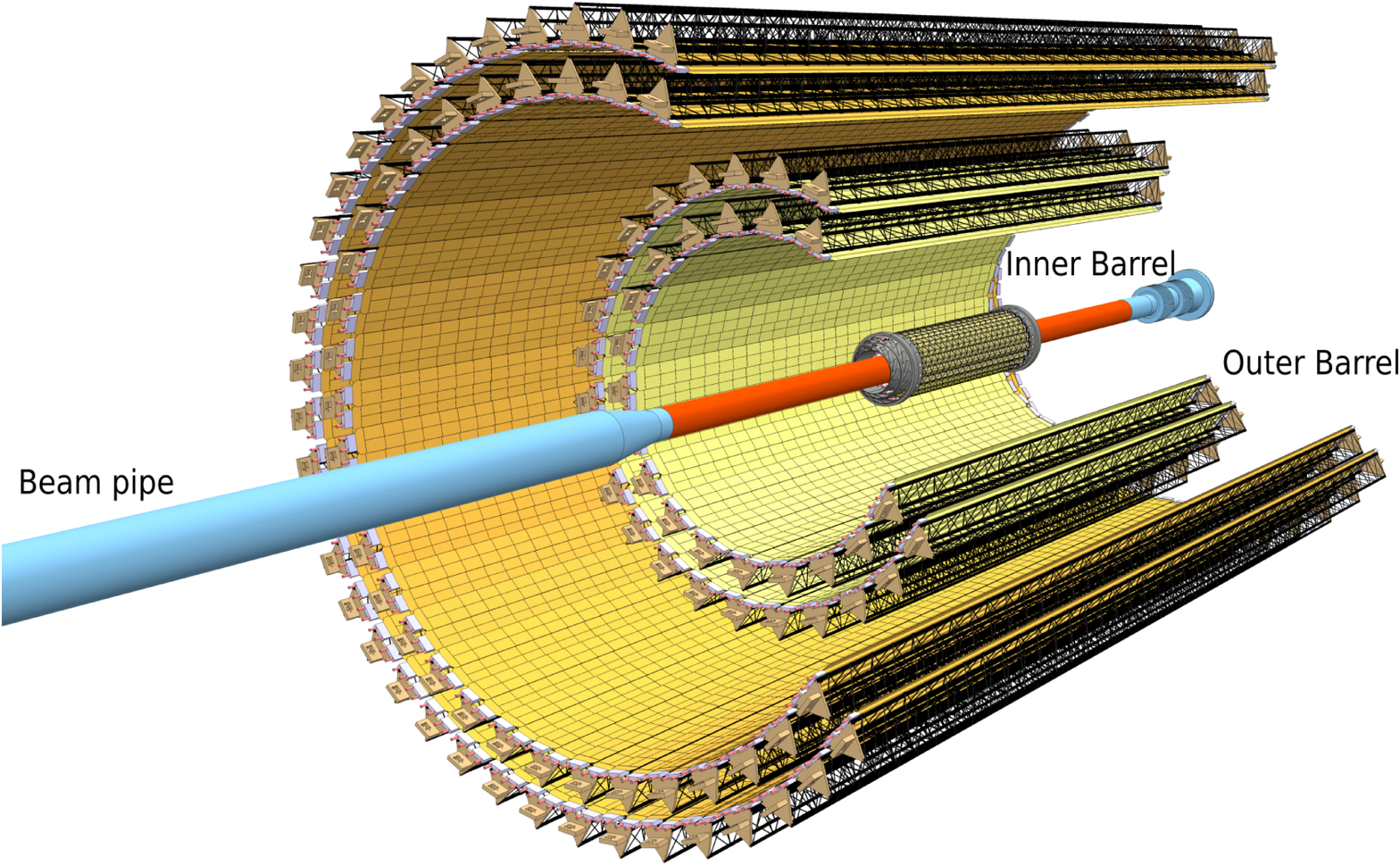}
\caption{Schematic layout of the upgraded ITS.}
\label{fig:itslyo}
\end{figure}

\begin{figure}
\centering
\includegraphics[width=0.67\textwidth]{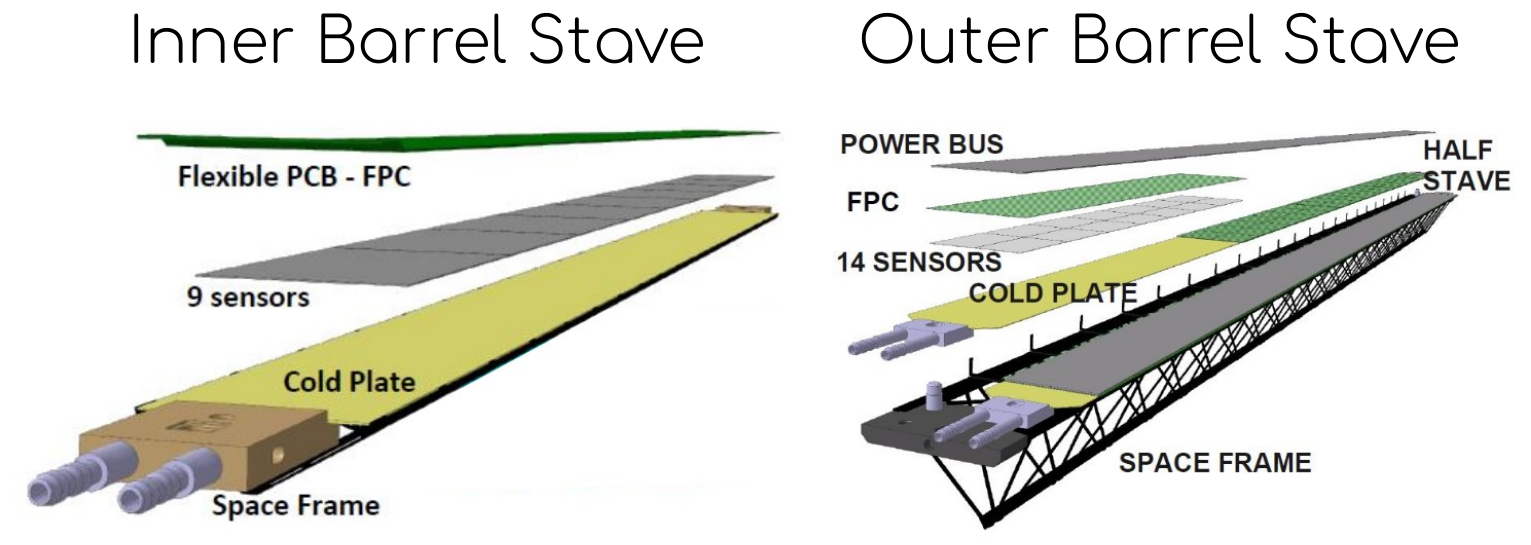}
\caption{View (not to scale) of an Inner Barrel Stave (left) and of an Outer Barrel Stave (right) of upgraded ALICE ITS.}
\label{fig:staves}
\end{figure}

To fullfil the requiments of the Run 3 physics program of the CERN Large Hadron Collider (LHC), the Inner Tracking System (ITS) of the ALICE experiment has been upgraded with seven cylindrical and concentric layers of silicon Monolithic Active Pixel Sensors (MAPS) called ALPIDE covering a total active area of about 10 m$^2$~\cite{ref:ITSTDR}. The detector is divided into three Inner Layers (IL), two Middle Layers (ML) and two Outer Layers (OL) as shown in Fig.~\ref{fig:itslyo}. The four outermost layers constitute the Outer Barrel (OB) while the three innermost layers form the Inner Barrel (IB). The main features of the upgraded ITS (simply called ITS2 in the following) compared to the ones of the older detector (called ITS1) are summarised in Table~\ref{tab:itsoldnew}.\\
The sensor, produced by TowerJazz in the 0.18 $\mu$m CMOS imaging process technology, features a pixel size of $\sim$27$\times$29 $\mu$m$^2$ with a total of 512 $\times$ 1024 pixels. The main characteristics of ALPIDE are:
\begin{enumerate}
\item a n-well diode about 100 times smaller than the pixel size, leading to a small capacitance of few fF;
\item a fast data driven encoder \cite{ref:pencoder};
\item a n-well of p-MOS transistors shielded by a deep p-well allowing a full CMOS circuitry within the active volume \cite{ref:aglieri,ref:sulic};
\item a pixel analogue signal amplified and digitized at a pixel level leading to a low power consumption of about 40 mW/cm$^2$;
\item the possibility to apply a moderate reverse bias voltage to the substrate between $-$6 V and 0 V.
\end{enumerate}
\begin{table}[t]
\caption{Main technological features of updraded ITS (ITS2) compared to the older ITS (ITS1). The radial positions are the average between the minimum and maximum radii of the innermost layer.}
  \label{tab:itsoldnew}
  \begin{tabular}{lcc}
  	\hline
  	& ITS1 & ITS2 \\
    \hline
      Readout rate      & up to 1 kHz  & $>$ 100 kHz (Pb--Pb) \\
      					&              & $>$ 400 kHz (pp)\\
      Material budget & 1.1\% X$_0$     & 0.36\% $X_0$ (Inner Barrel) \\
      				  &                  & 1.1\% $X_0$ (Outer Barrel)\\
      Pixel size      & 50$\times$425 $\mu$m$^2$  & 27$\times$29 $\mu$m$^2$ \\
      Pointing resolution ($p_{\rm T}$ = 500 MeV/$c$) & $\sim$240 $\mu$m ($z$) & $\sim$50 $\mu$m ($z$)\\
      								   & $\sim$120 $\mu$m ($r\varphi$) & $\sim$40 $\mu$m ($r\varphi$) \\
      Standalone tracking efficiency ($p_{\rm T}$ = 200 MeV/$c$) & $\sim$60\% & $\sim$90\% \\
      Radius (innermost layer) & 3.9 cm & 2.5 cm \\
    \hline
  \end{tabular}
\end{table}
In the ITS2, the ALPIDE chips are arranged in Hybrid Integrated Circuits (HIC) featuring 14 (9) ALPIDE chips for the OB (IB). A Flexible Printed Circuit (FPC) wire-bonded to the chips into a HIC is used for clock, control and data transmission towards and from the outside electronics \cite{ref:ITSTDR}. For IB chips, a readout at 1.2 Gb/s per chip is achieved while for the OB a group of 7 chips is readout with a single link at 400 Mb/s. The HICs are glued on a cold plate for chip cooling. The IB cold plate is attached to a ultra-light space-frame forming a Stave with a total length of 27 cm. In the OL(ML), 7(4) HICs are glued on cold plate to build an Half-Stave (HS) with a total length of 147(84) cm. Two HSs are then aligned on a carbon-fiber support structure (space frame) with $\sim$100 $\mu$m precision to assemble the final Stave. The Stave features also two power buses and two bias buses (one per HS) for chip powering and chip reverse bias voltage supply, respectively. A schematic view of the ITS2 Staves is given by Fig.~\ref{fig:staves}. In summary, ITS2 is composed by 192 Staves on which 24120 sensors are mounted for a total of 12.5 billion pixels, being the largest pixel tracker ever built.\\ 
%
\section{Data readout architecture}\label{sec:readout}
\begin{figure}
\centering
\includegraphics[width=0.77\textwidth]{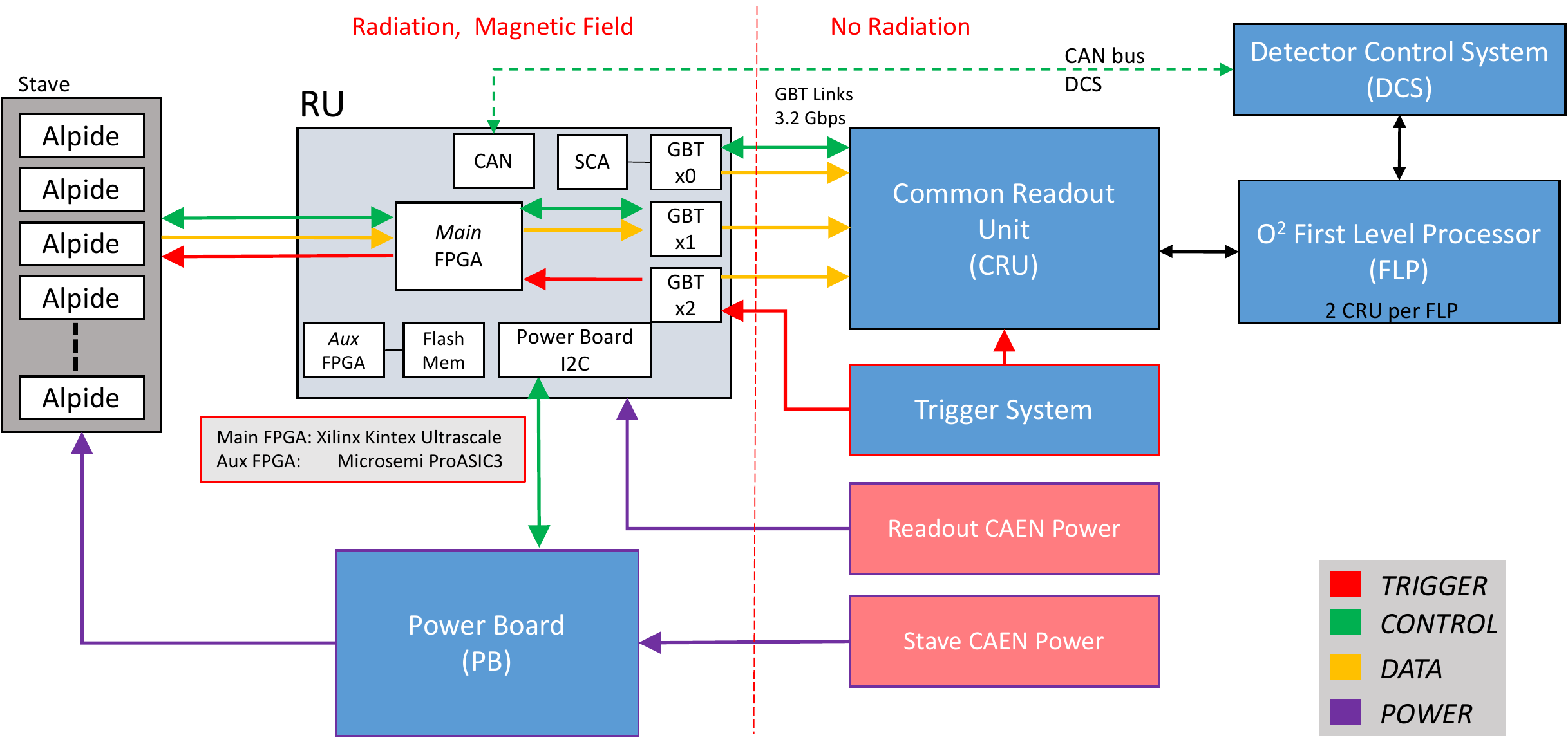}
\caption{Summary scheme of the data readout achitecture of ITS2.}
\label{fig:readout}
\end{figure}

The huge amount of channels of ITS is readout by 192 identical Readout Units (RU) which provide control and trigger, and read the high-speed data lines from the ALPIDE chips with a total data rate of up to $\sim$40 GB/s. To minimize the power consumption and achieve the maximum bit rate of 1.2 Gb/s, the RUs and the ITS2 Power Boards (PB) are located within the ALICE L3 magnet, in a magnetic field of 0.5 T, and are exposed to the radiation environment. Both sytems were tested and validated under these conditions \cite{ref:RUrad}. The RU design thus makes use of two programmable logic devices, one of them a flash-based FPGA, whereas the other is a Static RAM (SRAM)-based FPGA. The first is responsible for auxiliary tasks, such as configuring the other FPGA and periodically refreshing (“scrubbing”) its Configuration RAM (CRAM). The latter is implementing the main functionality of the RU, being in charge of data readout from the sensor, assembling sensor data into event frames and communicating those to the ALICE Offline and Online (O$^{2}$) computing system \cite{ref:o2}, receiving triggers from the ALICE Central Trigger Processor \cite{ref:ctp}, control, configuration and monitoring of the ALPIDE sensors, as well as control and monitoring of the PB, which provides the power to the ALPIDE. \\
The 192 RUs are connected through GigaBit Transceiver (GBT) links (up to 3 per RU) to 22 Common Readout Units (CRU) which are mounted into 13 First Level Processors (FLP, CPU servers) located about 200 m far from the detector, in a non-radiation environment, which are part of the O$^{2}$ DAQ farm. The CRUs are custom-FPGA based boards with up to 24 bi-directional fiber-optic links and a PCIe interface to the relative FLP. The FLPs communicate via a shared network with the processors of the Detector Control System (DCS) and transfer DCS data to and from the RUs via their associated CRUs. Trigger information is communicated from the trigger system to both the RUs and the CRUs so that event fragments from various RUs can be associated with the appropriate triggers. As a backup, when the O$^{2}$ system and its optical links are not available (e.g. during system maintenance) the RU is equipped with a CAN bus communication interface for control and monitoring.\\
The FLPs are connected to a farm of 250 Event Processing Nodes (EPN) shared among all the ALICE detectors which feature 32-core AMD Rome CPUs and 8 AMD MI50 graphics processing units (GPUs) each. By using the computing farm, the data can be processed synchronously, for time frame building, online reconstruction and calibration, or asynchronously, for final calibration and full reconstruction.\\
Finally, the EPN farm is connected to a 100 PB disk buffer. A 100 GB/s link connects the disk buffer to the permanent storage facility. Figure~\ref{fig:readout} summarizes the data readout chain of ITS2 up to the FLPs.
\section{Detector Control System and Data Quality Control}\label{sec:qcdcs}
As described in Sec.~\ref{sec:readout}, the FLPs communicate via a shared network with the machines of the Detector Control System and tranfer DCS data to and from the RUs via their associated CRUs. The DCS functionalities are accessible through a WinCC panel and they mainly include:
\begin{itemize}
\item a Finite State Machine (FSM) allowing for detector operations: powering and configuration. Different statuses of the detector are encoded by different colors for the various components of the detector, from the barrels to the single HICs; 
\item a full color-encoded (status-dependent) map of the detector with access to single stave functional parameters: power supply voltage and currents of the stave and its electronics, status of the power supply, temperature, status of the cooling lines;
\item widgets which allow to configure the detector for different kind of runs: physics, calibration, simulated data patterns.
\end{itemize}
An independent ITS Safety System (ITS2S) interlocks the power channels based on the stave temperatures and cooling loop statuses. \\
The DCS is used both by detector experts and central ALICE shifters for everyday operations with the detector (powering, configuration, calibration, etc.) and for the monitoring of all its functional parameters. Based on that, a set of alarms pops up in case one or more operating parameters overcome the limits of their optimal working ranges. \\
Another important branch in the context of the control systems is represented by the Data Quality Control System (simply called QC in the following) which is capable to synchronously monitor the following data:
\begin{itemize}
\item detector fake-hit rate: monitoring of detector raw hits and noisy pixels;
\item front-end electronics: monitoring of the number of channels in error and of the trigger flags;
\item decoding errors: monitoring of decoding errors for every chip;
\item clusters: monitoring of the cluster size and of the cluster occupancy in every detector sensor;
\item tracks: monitoring of track multiplicty in every readout frame, angular distributions ($\eta$ and $\varphi$), clusters per track, etc.;
\item calibration data: monitoring of calibration data during special calibration runs. The main variables are the amount of problematic and dead pixels in the detector and the in-pixel discriminator thresholds. This is extremely important so to guarantee a stable and efficient data taking. 
\end{itemize}
In addition, a set of post-processing tasks has been developed in order to monitor the stability of the above parameters over time on a run-by-run basis or during the run itself. As for the DCS, also the QC system is used on an everyday basis for the online quality control of the data but also for the offline reconstruction in order to ensure the preparation of a good data sample for the subsequent physics analyses. \\
Finally, the quality control of the data happens also at Monte Carlo level to check the tracking efficiency, the impact-parameter distributions and the raw particle momentum distributions. Every QC task is developped in C++ and is part of the ALICE O$^{2}$ software \cite{ref:qc}. 
%
\section{Detector commissioning, installation and preparation for Run 3}\label{sec:commissioning}
\begin{figure}
\centering
\includegraphics[width=0.67\textwidth]{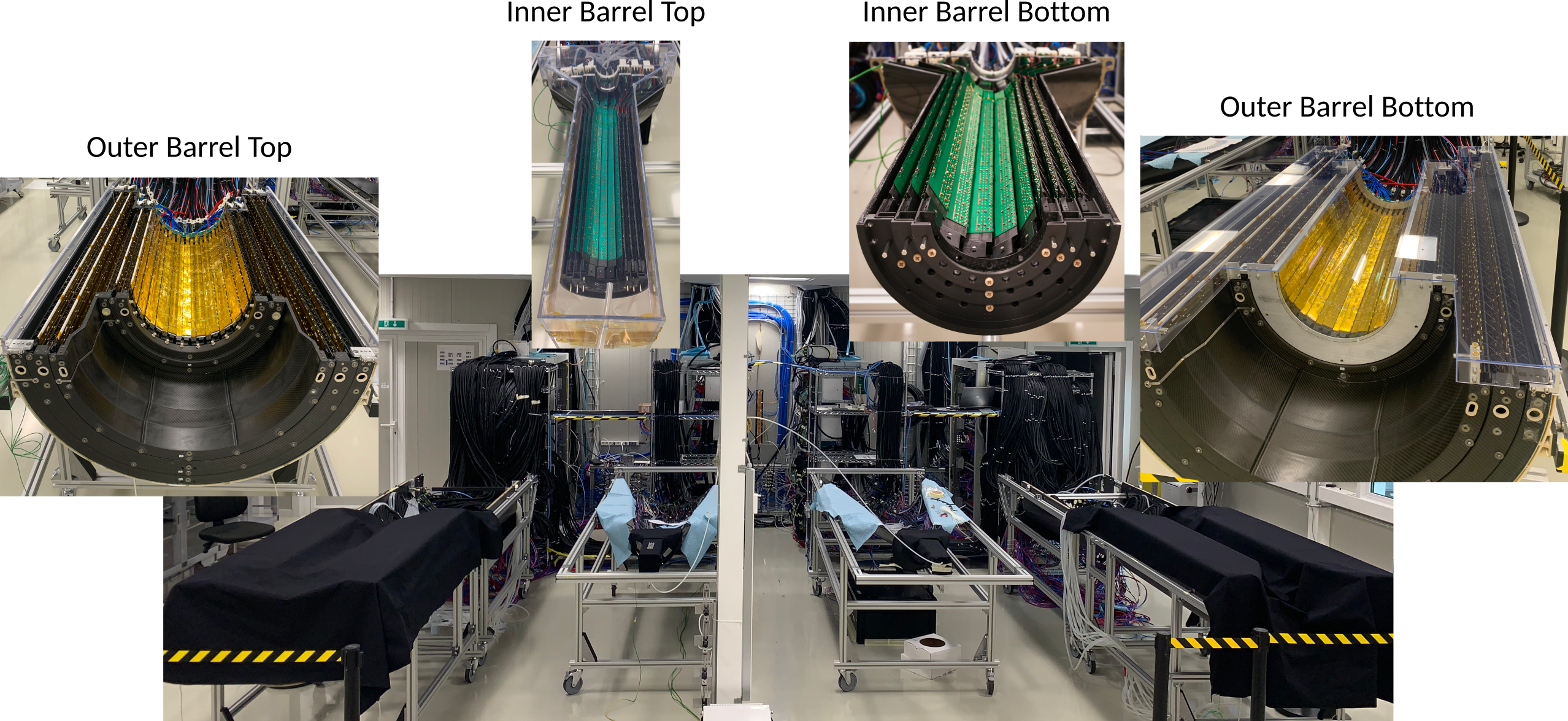}
\caption{The ALICE ITS2 in the CERN clean room split in half-barrels during the on-surface commissioning.}
\label{fig:surface}
\end{figure}

\begin{figure}
\centering
\includegraphics[width=0.77\textwidth]{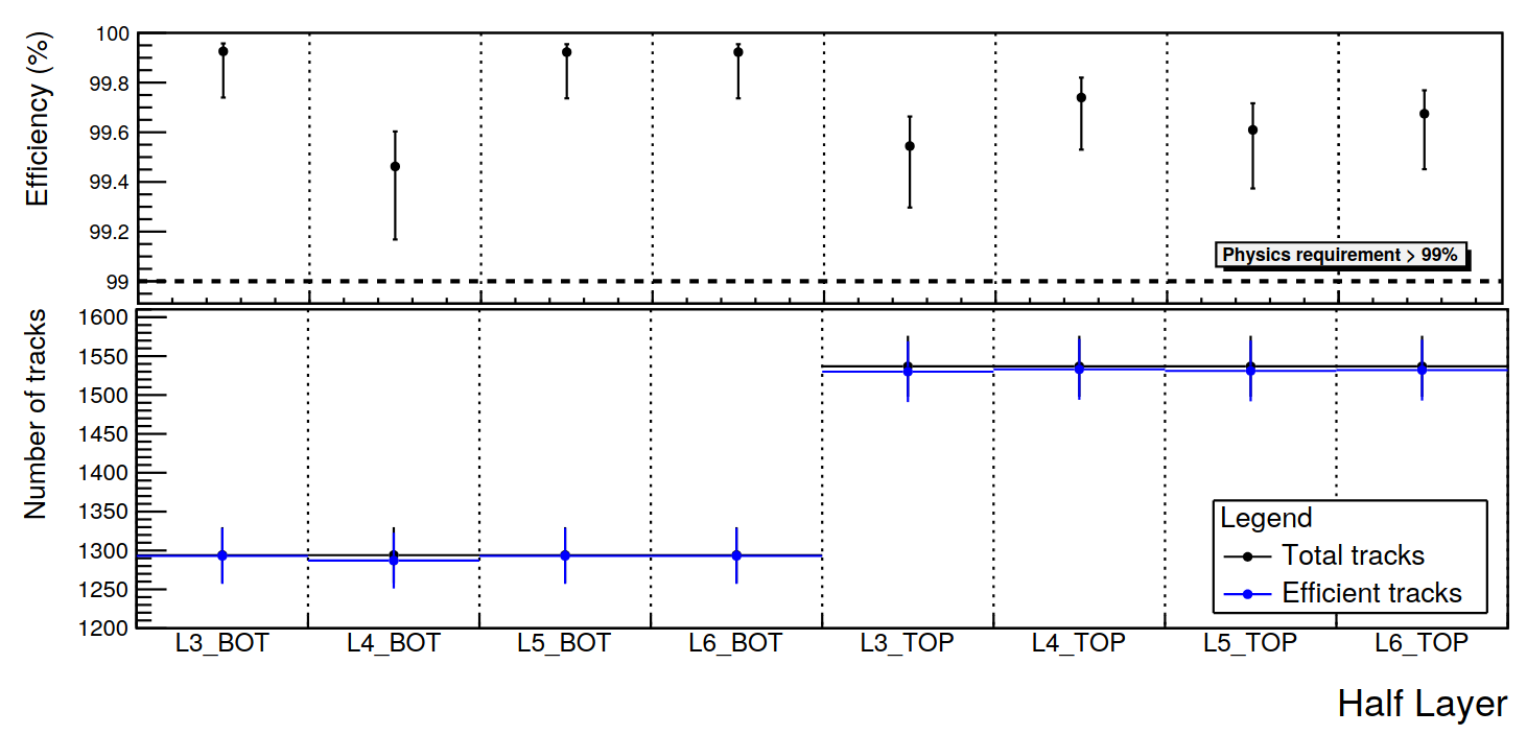}
\caption{Detection efficienty of ITS2 outer half-barrels calculated with cosmic muons during the on-surface commissioning. The bottom panel shows the number of cosmic tracks considered for the analysis in each half-barrel.}
\label{fig:eff}
\end{figure}

\begin{figure}
\centering
\includegraphics[width=0.67\textwidth]{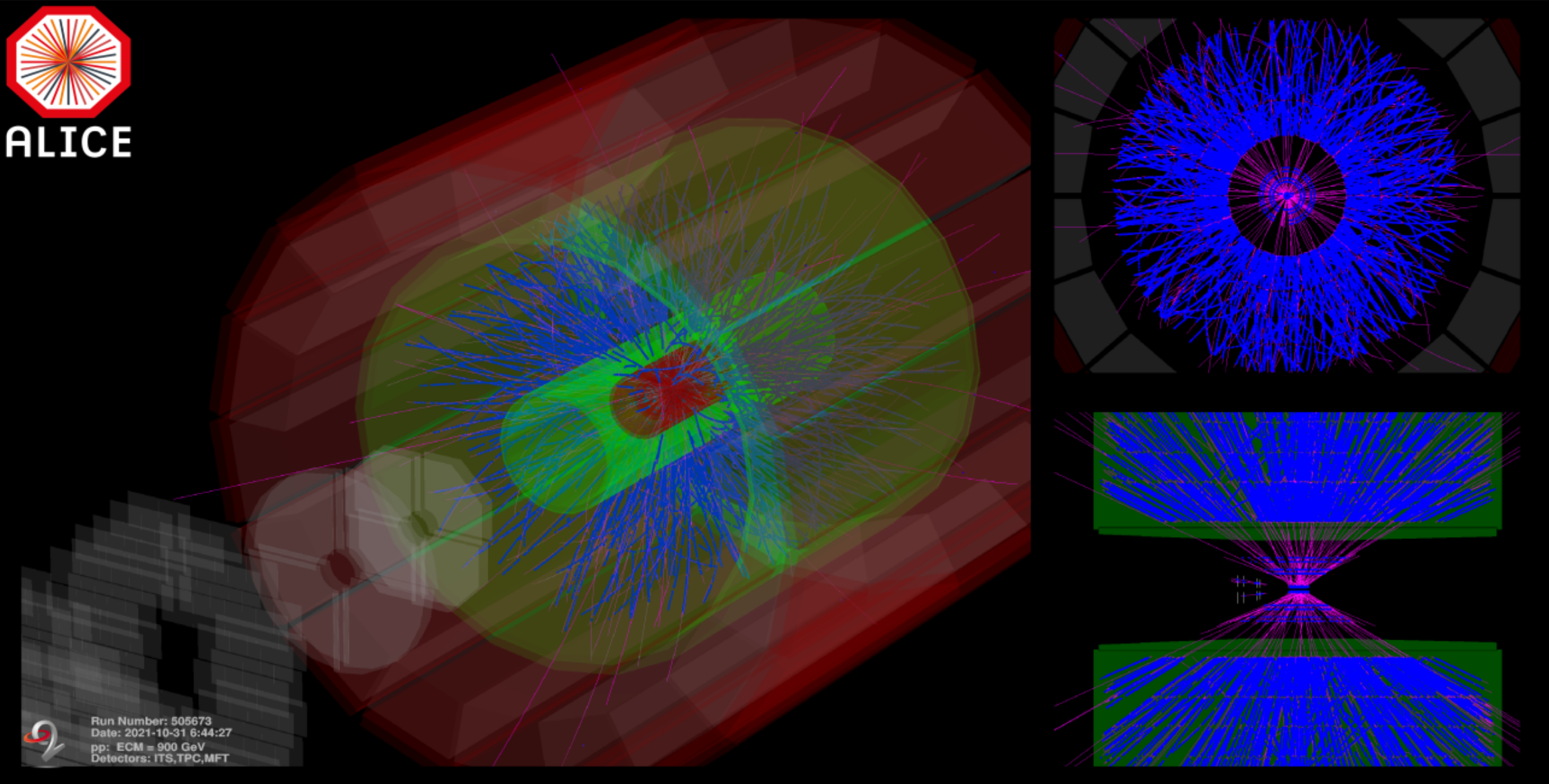}
\caption{Event display of the very first pp pilot collisions at $\sqrt{s}$ = 900 GeV of October 2021.}
\label{fig:evepilot}
\end{figure}

The detector assembly was performed in a clean room, where staves have been mounted in two IB and two OB half-barrels, connected to final services (power, cooling, readout) installed in an adjacent room (Fig.~\ref{fig:surface}).
The on-surface commissioning of the detector started in September 2019 and continued until December 2020 to validate in sequence each of the four half-barrels after assembly.
In the last phase of the on-surface commissioning the detector was continuosly in operation with 24/7 shifts: several cosmic and calibration runs were recorded and analysed with the aim of validating the detector stability. An important result of this phase is shown in Fig.~\ref{fig:eff} where the average detection efficiency of the outermost half-barrels is plotted. As can be seen, an efficiency above 99.5\% is obtained. This is well in line with the detector requirements for Run 3. The bottom panel of the plot shows the number of cosmic tracks taken into account for the measurements, different for top and bottom barrels due to their separated position in the laboratory.\\
The installation of the detector in the cavern has been performed in two phases: first the outer barrel was installed inside the ALICE Time Projection Chamber (TPC) and, successively, the inner barrel was positioned around the beampipe. The installation of the full tracker represented a challenge mainly because of a small clearance between adjacent staves of only 1.2 mm (when approaching the two half-barrels) and of a clearance around the beam pipe of only 2 mm. In addition, the manipulation of the barrels had to be performed from a distance of about 4 m. Several cameras where mounted in different spots to constantly monitor the position of the barrels during the insertion and the final approach of the two halves. The installation of the full detector was completed in May 2021. A basic verification of the detector followed after installation including power tests, readout tests and resistance measurements of the data lines to ensure the reliability of all cable connections. \\
The beginning of summer 2021, till July, was dedicated to the ITS2 standalone commissioning with 24/7 shifts. In this phase the detector was fully recommissioned with continuos cosmic data taking and calibration runs. In addition, the full readout chain has been tested by simulating the data load provided by high multiplicity pp collisions and central lead--lead collisions: the hits extracted from the Monte Carlo simulation, were used to pulse the pixels involved so to simulate the hit occupancy of real collisions. \\
In July 2021 the ALICE global commissioning started with the aim of validating the full ALICE software for data taking and reconstruction in view of the very first pp pilot collisions at LHC injection center-of-mass energy ($\sqrt{s}$ = 900 GeV) which happened at the end of October 2021. Figure~\ref{fig:evepilot} shows an event display of that period with reconstructed tracks in ITS2 and in the external TPC in one Time Frame (TF) corresponding to a time interval of 11.5 ms containing about 6 collisions (due to the low interaction rate (IR) of the bunches). 
\section{Calibration of ITS2}\label{sec:calib}
\begin{figure}
\centering
\includegraphics[width=0.87\textwidth]{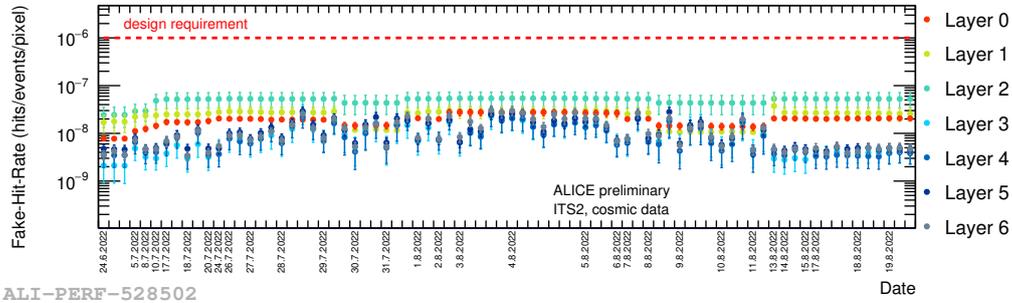}
\caption{Average fake-hit rate for each layer of ITS2 as a function of time after masking about 0.015\% of the pixels of the full detector.}
\label{fig:fhrstability}
\end{figure}

\begin{figure}
\centering
\includegraphics[width=0.8\textwidth]{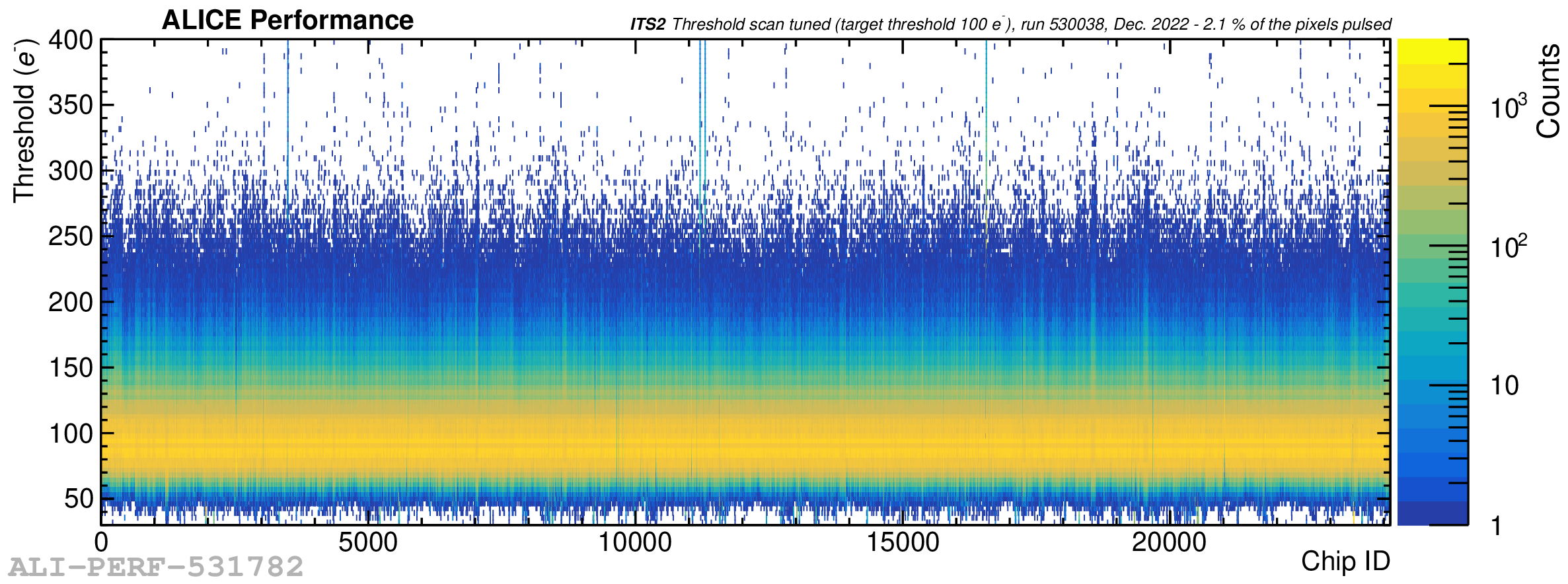}
\caption{Bi-dimensional distribution of the threshold of 2\% of the pixels in all the 24120 sensors of ITS2 after a tuning with a target threshold of 100 electrons. About 2.1\% of the pixels are considered for this measurement, which is enough to evaluate the quality of the detector calibration.}
\label{fig:tune}
\end{figure}

The calibration of ITS2 mainly consists in the determination and tuning of its in-pixel discriminator thresholds, and in the masking of its noisy pixels based on the selected target threshold. The measurement of the in-pixel threshold is based on the injection of an analogue pulse with amplitudes ranging from 0 to 500 electrons. The pixel response is fit with an error function to extract the 50\% point (threshold) and the sigma (ENC noise). Similarly, the tuning of the thresholds is obtained by injecting a fixed pulse equal to the target threshold (typically 100 or 150 electrons at which the ALPIDE is known to be fully efficient \cite{ref:sulic}), and by varying the registers which are responsible for the variation of the threshold itself. \\
Once the threshold is fixed, a standard cosmic run is used to tag the most noisy pixels in the detector and to prepare a detector configuration containing the proper masking: every pixel in the IB(OB) with an occupancy above 10$^{-2}$(10$^{-6}$) hits/event is tagged as noisy and masked. The looser selection for the IB is chosen in order to maximize the detection efficiency. With a typical average threshold of 100 electrons, the number of masked pixels amounts to only $\sim$0.015\% of the total. This leads to an overall fake-hit rate of about 10$^{-8}$ hits/event/pixel in average (stable over time), well below the required 10$^{-6}$ hits/event/pixel by design as shown in Fig.~\ref{fig:fhrstability}.
In order to run the calibration scans, the ITS2 RUs contain a sequencer to trigger the pulses to the chips, which allows the scan execution to be controlled entirely by the DCS software.\\
The main challenge resides in the decoding and the threshold extraction of the 12.5 billion channels.  A threshold scan of the full detector with 50 charge injection points and 50 hits per point results in about 3 $\times 10^{13}$ hits or approximately 100 TB of raw hit data. If the scan is performed as fast as the on-detector bandwidth allows, this data is collected in slightly less than 1 hour, resulting in a data rate of 20--30 GB/s. However, a full scan is generally used as a reference and it is not needed on a daily basis. In fact, to ensure a good threshold calibration of the detector it is sufficient to pulse about 1--2\% of the pixels with the advantage of completing the scan in less than 5 minutes. \\
In order to cope with the amount of data the analysis is performed in a distributed manner on the EPNs. A special configuration of the sub-timeframe builder is used, since the requirements of the calibration are orthogonal to standard data taking: whereas in data taking single events of the full detector are assembled to one time-frame and shipped to available EPNs, in calibration a number of consecutive events belonging to a group of chips (a so-called ``link") must be analysed together on the same EPN. As a consequence, the sub-timeframe builder sends all events of a detector link to the same processing node, thus reducing significantly the bandwidth used for data transfer between the processing nodes. Figure~\ref{fig:tune} shows the distribution of the threshold for 2\% of the pixels in all the 24120 sensors of ITS2 after a tuning with a target threshold of 100 electrons. The typical ENC noise is of only 5 electrons and it is observed to be stable for all chips. 
In summary, every parameter related to calibration (addresses of the pixels to be masked, ALPIDE register settings for threshold tuning) is used to prepare a new detector configuration in the configuration database by using a dedicated WinCC panel in the DCS sofware. 
\section{Start of LHC Run 3 and ITS2 performance}\label{sec:commissioning}
\begin{figure}
\centering
\includegraphics[width=0.67\textwidth]{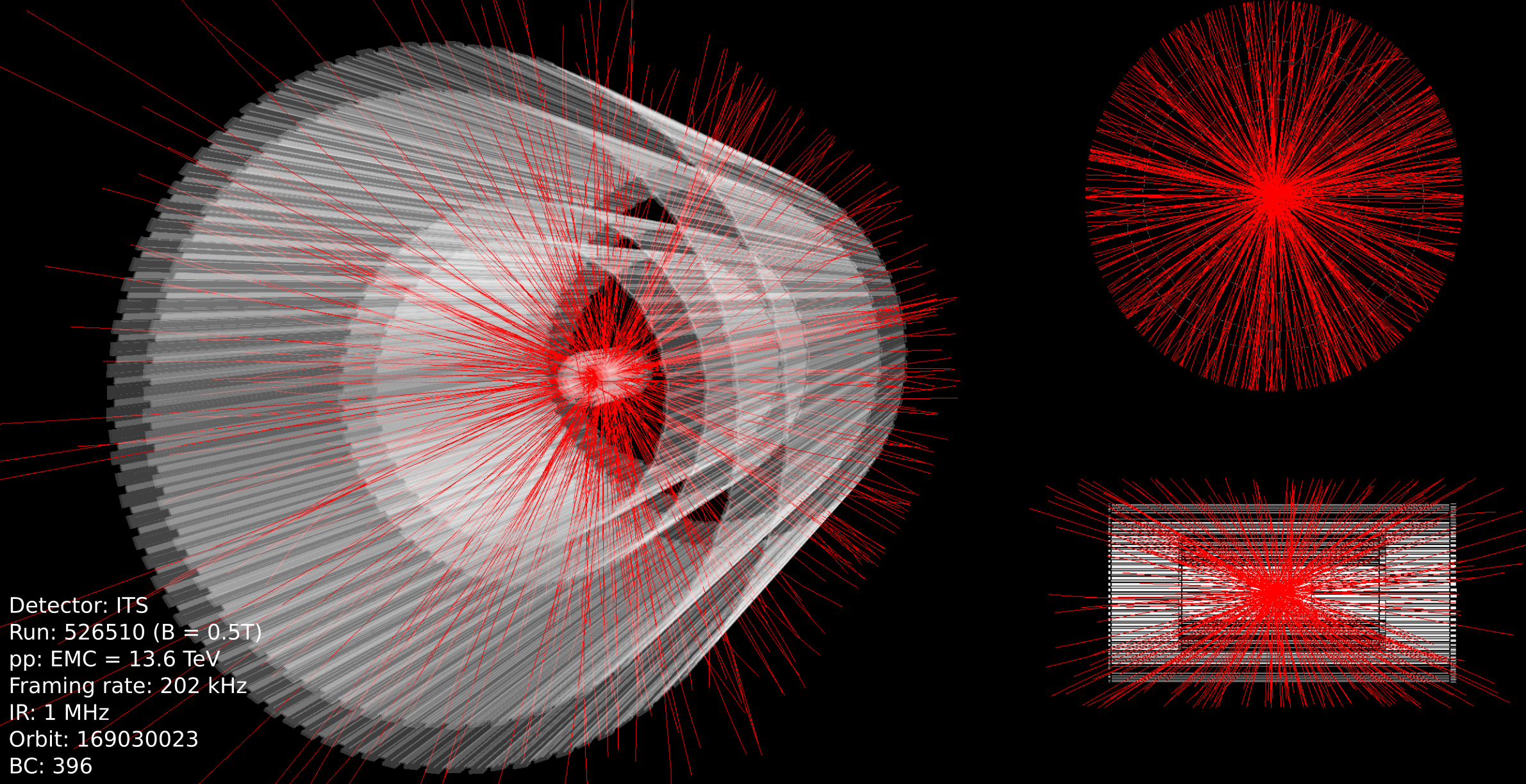}
\caption{Event display of 1 readout frame in pp collisions at $\sqrt{s}$ = 13.6 TeV with an average IR of the proton bunches of about 1 MHz. The ITS2 framing rate was 202 kHz which corresponds to a time-length of the frame of about 5 $\mu$s.}
\label{fig:everun3}
\end{figure}

\sloppy
The LHC Run 3 started officially on July 5th, 2022 with pp collisions at the unprecedented energy of $\sqrt{s}$ = 13.6 TeV. As of today, the ALICE detector accumulated about 700h of data taking with pp collisions and 10.6 hours of Pb--Pb pilot collisions at a center-of-mass energy per nucleon pairs of $\sqrt{s_{\rm NN}}$ = 5.36 TeV. The nominal IR of the proton bunches in LHC is about 500 kHz and the ITS2 is readout in continuos mode at a framing rate nominally set to 202 kHz. High-rate rate scans have also been performed by increasing the IR up to 4--5 MHz. For the first Pb--Pb test, the IR was about 42 Hz and the ITS2 took data in continuous mode at a framing rate of 67 kHz. The loss of acceptance of ITS2 during the runs is generally below 1\% which is in agreement with the design requirements for Run 3. At every beam dump, a scan of the pixel thresholds is performed to ensure that the detector remains fully efficient in the subquent runs.\\
Figure~\ref{fig:everun3} shows an event display with reconstructed ITS2-standalone tracks in pp collisions at \mbox{$\sqrt{s}$ = 13.6 TeV} at an IR of the proton bunches of about 1 MHz.
\subsection{Performance of ITS2}
\begin{figure}
\centering
\includegraphics[width=0.55\textwidth]{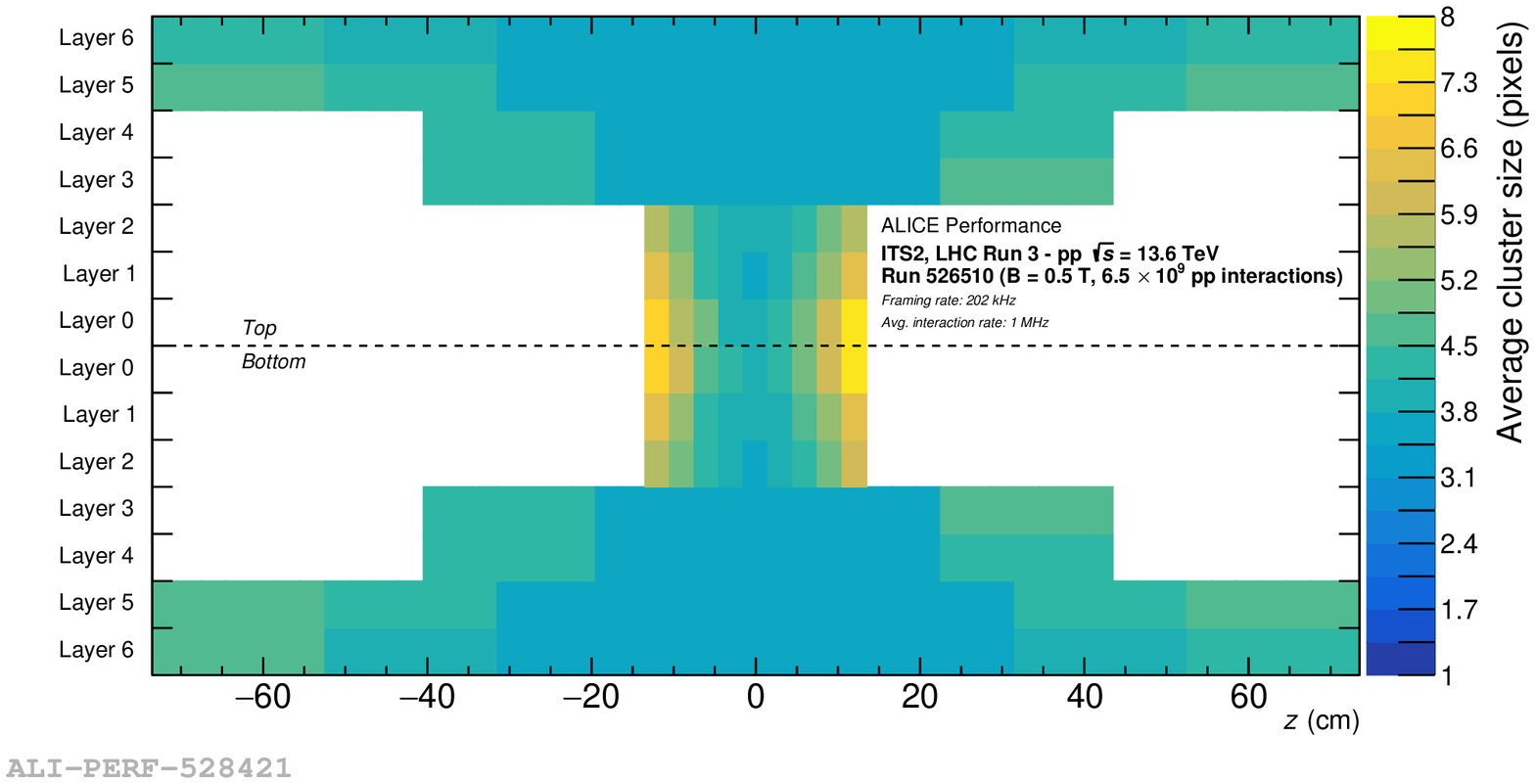}
\hfill
\includegraphics[width=0.4\textwidth]{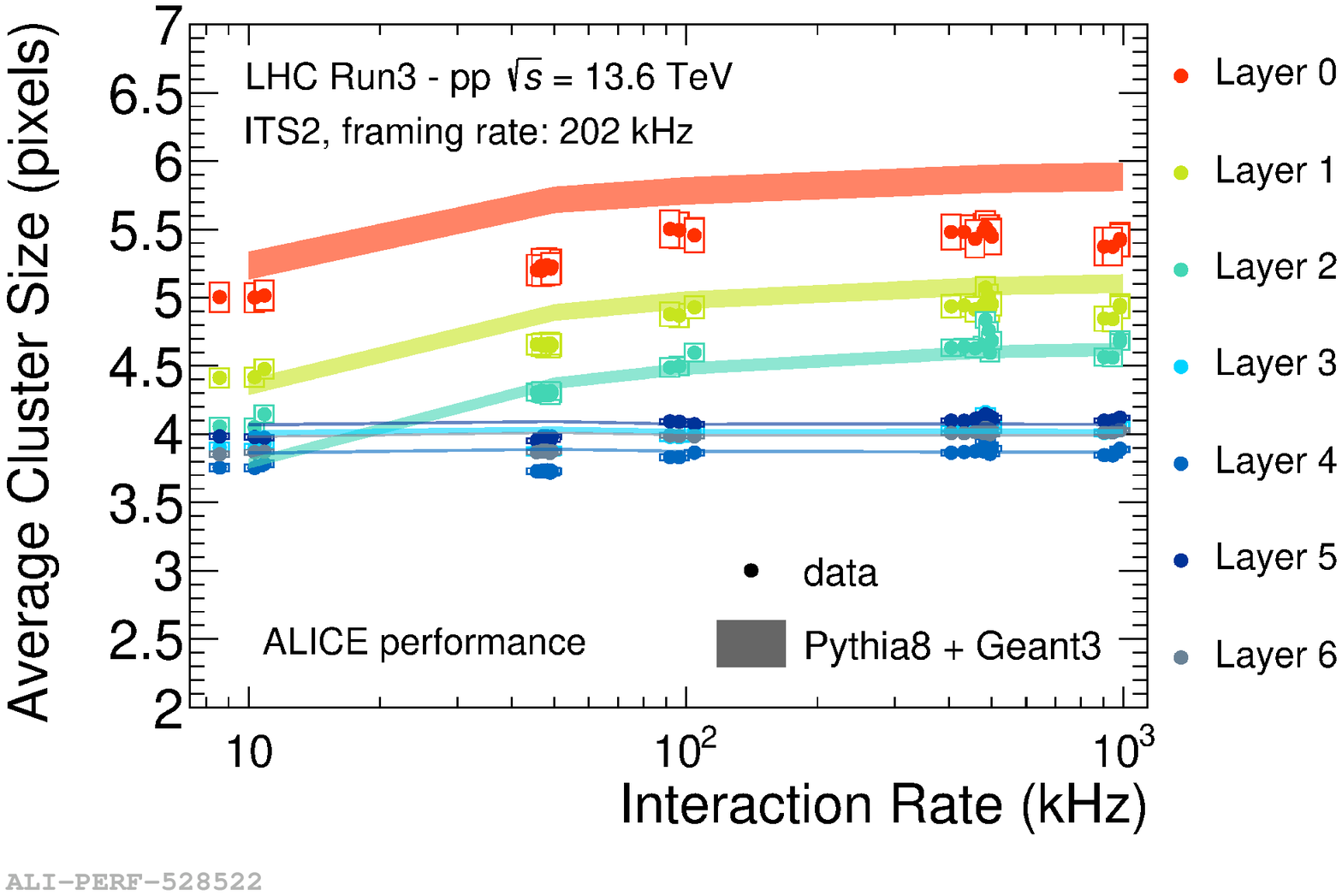}
\caption{Left figure shows the average cluster size for each half-layer of the ITS2 in a pp run of Run 3 ($\sqrt{s}$ = 13.6 TeV) at an IR of the bunches of about 1 MHz and at an ITS2 framing rate of 202 kHz. The minimum granularity for the layers 0, 1 and 2 is the single chip while for the outermost later it is given by the single HIC. Right figure shows the dependence of the cluster size on the bunch IR including the results from a simulation based on Geant 3 and Pythia 8.}
\label{fig:cls}
\end{figure}

\begin{figure}
\centering
\includegraphics[width=0.55\textwidth]{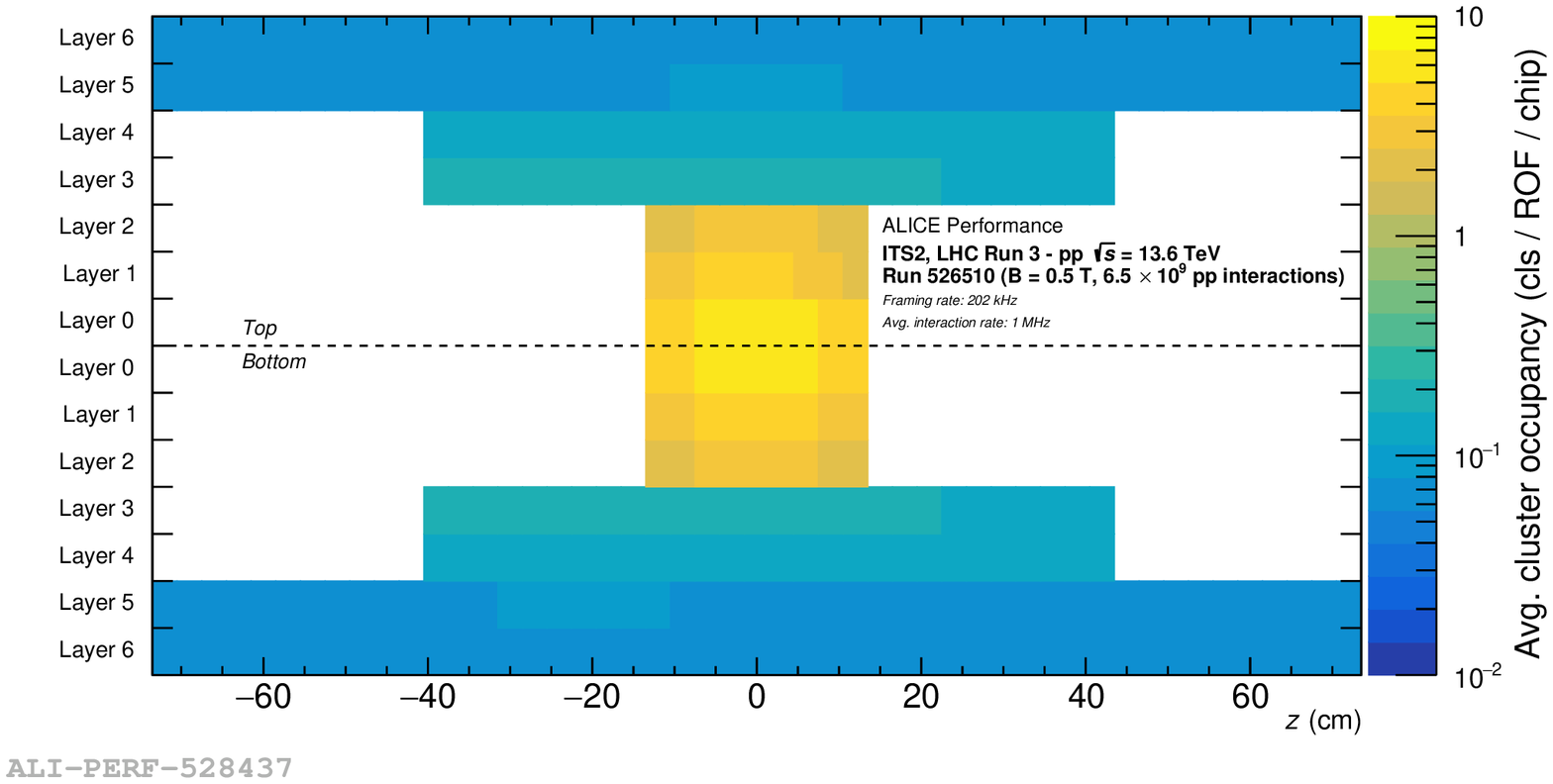}
\hfill
\includegraphics[width=0.4\textwidth]{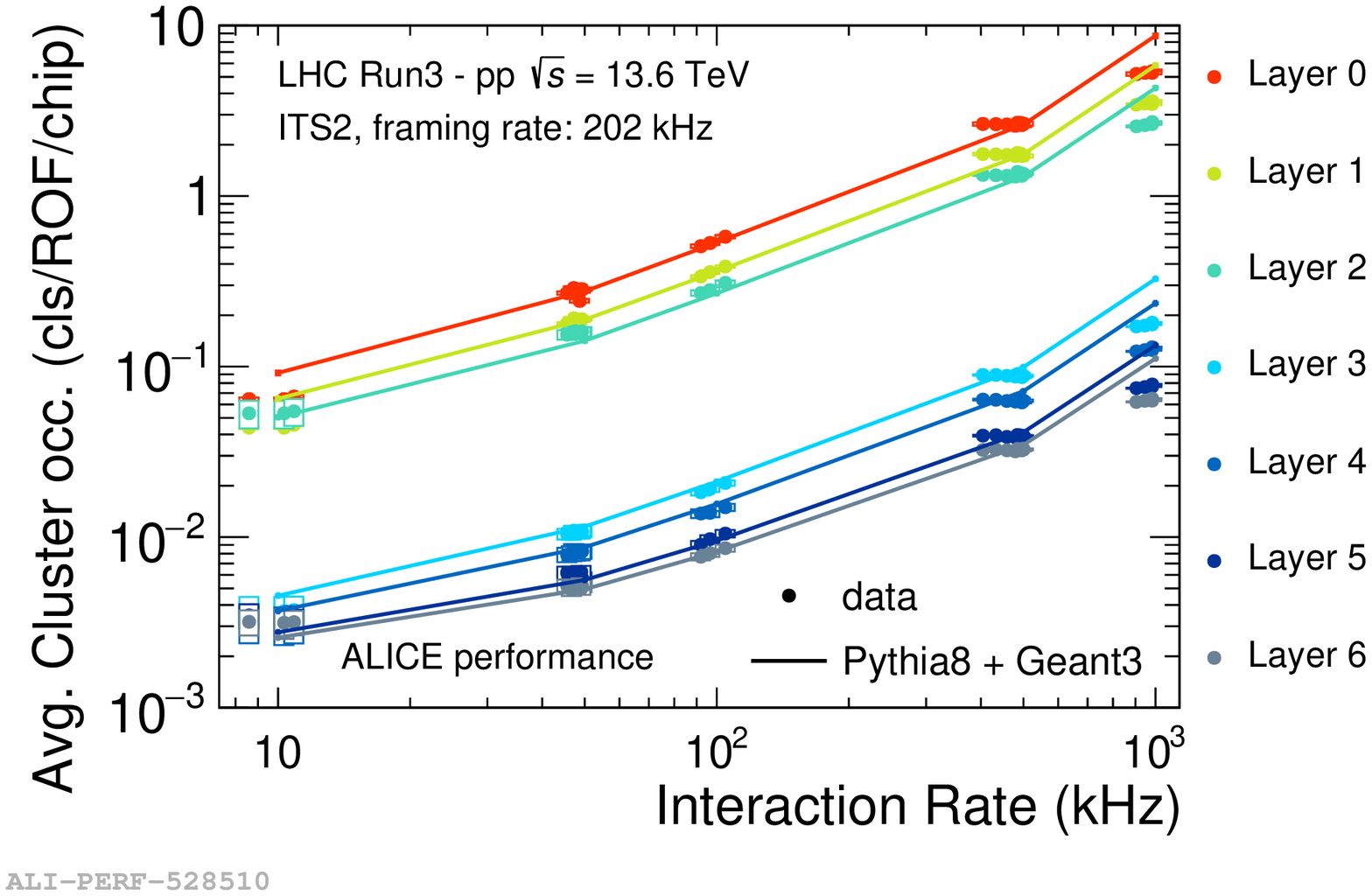}
\caption{Left figure shows the average cluster occupancy (number of clusters per readout frame per chip) for each half-layer of the ITS2 in a pp run of Run 3 ($\sqrt{s}$ = 13.6 TeV) at an IR of the bunches of about 1 MHz and at an ITS2 framing rate of 202 kHz. The minimum granularity for the layers 0, 1 and 2 is the single chip while for the outermost later it is given by the single HIC. Right figure shows the dependence of the cluster occupancy on the bunch IR including the results from a simulation based on Geant 3 and Pythia 8. See text for further details.}
\label{fig:clsocc}
\end{figure}

\begin{figure}
\centering
\includegraphics[width=0.5\textwidth]{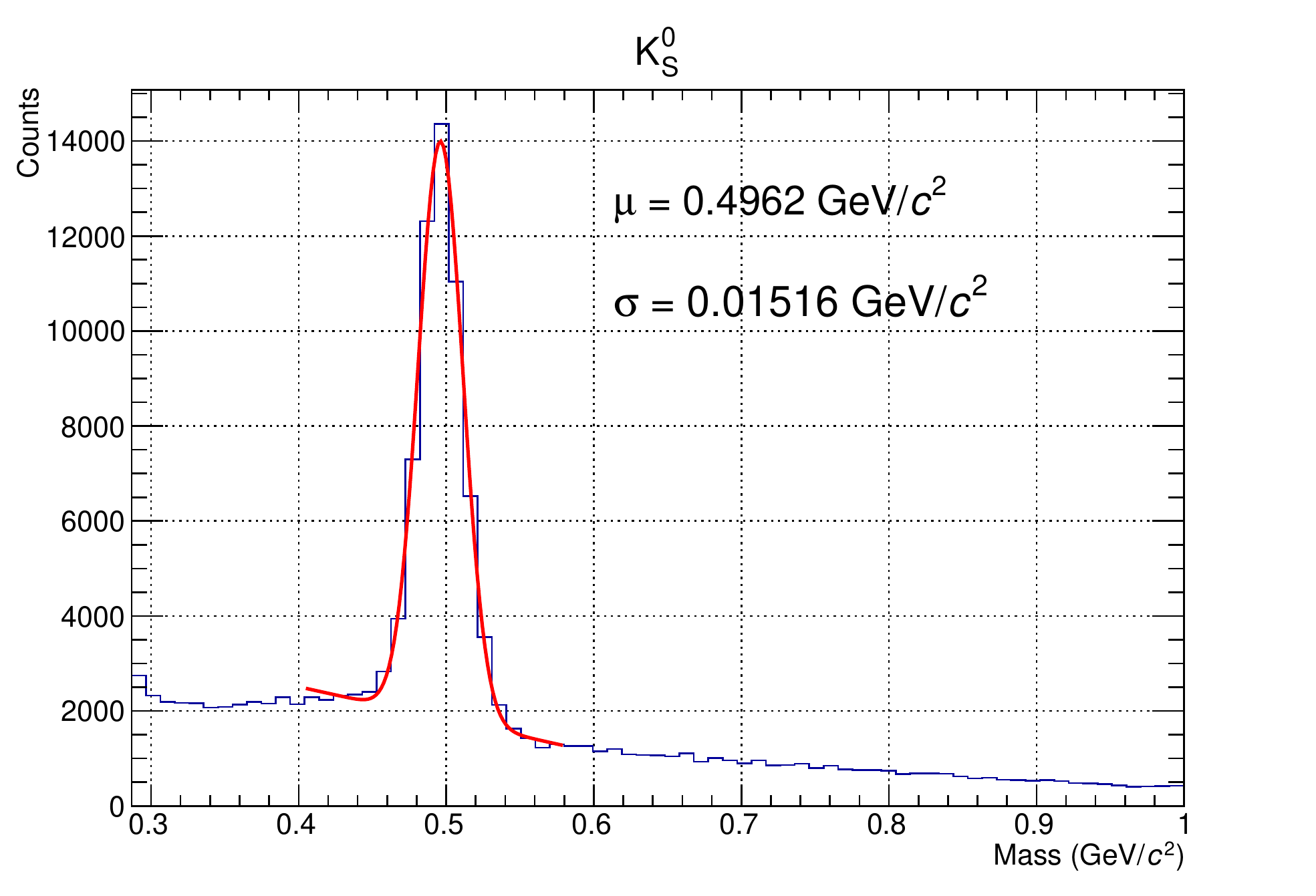}
\caption{$p_{T}$-integrated invariant mass peak of the K$^{0}_{S}$ obtained with the ITS-standalone tracking in pp collisions at $\sqrt{s}$ = 13.6 TeV.}
\label{fig:k0s}
\end{figure}

\begin{figure}
\centering
\includegraphics[width=0.47\textwidth]{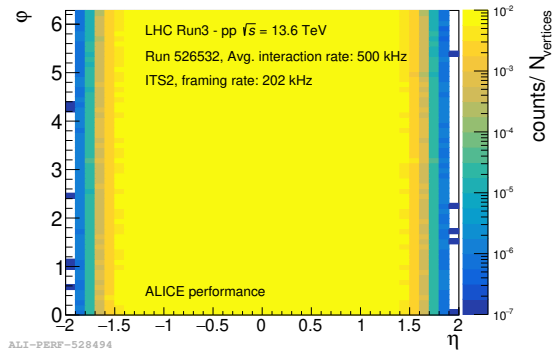}
\caption{Correlation between the azimuthal angle $\varphi$ and the pseudo-rapidity $\eta$ of the reconstructed ITS2 tracks in pp collisions at $\sqrt{s}$ = 13.6 TeV at an IR of about 1 MHz.}
\label{fig:angdist}
\end{figure}

\sloppy
This section aims at illustrating the first performance results of ITS2 in Run 3 pp collisions at $\sqrt{s}$ = 13.6 TeV. The first piece of information which is important for a pixel detector is the size of the clusters and their occupancy. Studies with machine learning techniques are ongoing in order to exploit the cluster size for the identification of pions, kaons and protons down to an unprecedented low transverse momentum $p_{\rm T}$ $<$ 0.1 GeV/$c$. Left panel of Fig.~\ref{fig:cls} shows the average cluster size for each half-layer of the ITS2 in a pp run of Run 3 ($\sqrt{s}$ = 13.6 TeV) at an IR of the bunches of about 1 MHz and at an ITS2 framing rate of 202 kHz. The minimum granularity for layers 0, 1 and 2 is the single chip while for the outermost layers it is given by the single HIC. As shown in Fig.~\ref{fig:cls}, the cluster size ranges between 3 and 8 pixels depending on the pseudo-rapidity $\eta$. The dependence of the cluster size on the bunch IR is shown in the right panel of Fig.~\ref{fig:cls} including the results from a simulation based on 20000 events generated with Pythia 8 \cite{ref:pythia8} and propagated with Geant 3 \cite{ref:geant3} through the detector material. Based on what shown in Fig.~\ref{fig:fhrstability}, the noise in the simulation is set to 2 $\times$ 10$^{-8}$ (3 $\times$ 10$^{-9}$) hits/event/pixel for the IB(OB). In general a good agreement between data and simulation is observed in average, within 5-6\%. The main source of discrepancy resides in the approximation of the noise simulation where only an average per barrel is considered instead of a single-chip value. The slight dependence of the cluster size on the IR can be explained by the fact that at low IR, the noise hits (1-pixel clusters) become comparable to the physics hits ($>$ 1-pixel clusters) lowering the cluster size as a consequence.\\
The average amount of clusters per readout frame and per chip in the full detector is instead shown in the left part of Fig.~\ref{fig:clsocc} for a pp run of Run 3 ($\sqrt{s}$ = 13.6 TeV) at an IR of the bunches of about 1 MHz and at an ITS2 framing rate of 202 kHz. With these parameters, the cluster occupancy ranges between 0.1 and 10 clusters per readout frame per chip depending on $\eta$. As opposed to the cluster size, the cluster occupancy is strongly dependent on the IR as shown on the right panel of Fig.~\ref{fig:clsocc} together with the very same simulation results explained at the beginning of this paragraph. In general a good agreement between data and simulation is observed considering the limitations previously explained.\\
The tracks in the ITS2 are recontructed by using a Cellular Automaton (CA) algorithm synchronously running on the EPNs. After a preliminary vertex position estimation, needed as seed, the tracking phase is constituted by three macro steps: tracklet finding, cell finding and track finding \cite{ref:ITSTDR}. In a subsquent phase, the ITS-standalone tracks are matched with the TPC tracks as explained in detail in Ref.~\cite{ref:tpctrack}. The tracking algorithm performance is strongly influenced by the detector alignment (obtained with the Millepede software \cite{ref:millepede}) which is currently in its close-to-final version. At the moment, at $p_{\rm T}$ $>$ 3 GeV/$c$ a pointing resolution of about 10 $\mu$m is measured (residual misalignment contributions are not negligible here) in $r\varphi$ direction while at lower momenta the resolution agrees with design specifications (see Table~\ref{tab:itsoldnew}). In addition, the signal of the K$^{0}_{S}$ can be already extracted as shown in Fig.~\ref{fig:k0s} with a preliminary study using ITS-standalone tracks in Run 3 pp collisions at \mbox{$\sqrt{s}$ = 13.6 TeV}. Further studies are ongoing to fine tune the detector alignment parameters and to study the performance of the detector in different conditions.\\
Finally, the angular distribution (correlation between the azimuthal angle $\varphi$ and the pseudo-rapidity $\eta$) of the reconstructed ITS2 tracks is shown in Fig.~\ref{fig:angdist} for pp collisions at $\sqrt{s}$ = 13.6 TeV at an IR of about 1 MHz. 
%
\section{Summary}
The upgraded Inner Tracking System (ITS2) of the ALICE experiment is composed by seven layers of silicon Monolithic Active Pixel Sensors with a total number of 12.5 billion pixels, being the largest pixel tracker ever built. It has been successfully installed in the ALICE cavern in May 2021 after about 1.5 years of commissioning on surface. The detector is readout by 192 identical FPGA-based readout units which communicate with the ALICE Offline and Online farm made of 13 First Level Processors (for ITS2 only) and 250 Event Processing Nodes for the synchronous and asynchronous calibration and reconstruction of the data. The Detector and Data Quality Control Systems are available to monitor the detector and its services, to operate the detector for data taking and to synchronously check the quality of the data. \\
The calibration of the detector mainly consists in the tuning of the in-pixel discriminator thresholds, in their measurement and in the monitoring of their stability over time. The calibration of the detector channels is challenging due to its large number channels. As a consequence, the procedure fully exploits the parallel processing on the powerful Event Processing Nodes.\\
The commissioning in the cavern, completed in June 2022, allowed the resolution of various hardware issues related to services and the full validation of the software for data taking and reconstruction. Since the start of the LHC Run 3 in July 2022, ALICE accumulated about 700h of data taking and the ITS2 main functional parameters were observed to be stable over time: average fake-hit rate at a level of 10$^{-8}$ hits/event/pixel by masking only $\sim$0.015\% of the total number of pixels, an average cluster size between 3 and 8 pixels and a realiable track reconstruction and detector alignment in software. ITS2 will continue to record data for the full Run 3 and it is ready for the first lead--lead collisions at top energy foreseen in October 2023.

\end{document}